\documentclass[runningheads,a4paper]{llncs}
\usepackage[cp1251]{inputenc}
\usepackage[english]{babel}
\usepackage{amssymb,amsmath}
\usepackage{textcomp}
\usepackage{graphicx}
\usepackage[all]{xy}
\usepackage[usenames]{color}
\usepackage{cite}
\usepackage{dsfont}

\begin{document}

\pagestyle{headings}  

\mainmatter              
\title{Secure Generators of $q$-valued Pseudo-Random Sequences on Arithmetic Polynomials}
\titlerunning{Secure generators of $q$-valued pseudo-random sequences}  
%
\author{Oleg Finko\inst{1} \and Sergey Dichenko\inst{1} \and Dmitry Samoylenko\inst{2} }
\authorrunning{Oleg Finko et al.} 
%
\tocauthor{Sergey Dichenko}
\institute{Institute of Computer Systems and Information Security of Kuban State Technological University, Krasnodar, Moskovskaya St., 2, 350072, Russia\\
\email{ofinko@yandex.ru}\and
Mozhaiskii Military Space Academy, St. Petersburg, \\
Zhdanovskaya St., 13, 197198, Russia\\
\email{19sam@mail.ru}
}
\maketitle
\begin{abstract}
A technique for controlling errors in the functioning of nodes for the formation of $q$-valued
pseudo-random sequences (PRS) operating under both random errors and errors generated
through intentional attack by an attacker is provided, in which systems of characteristic
equations are realized by arithmetic polynomials that allow the calculation process
to be parallelized and, in turn, allow the use of redundant modular codes device.

\end{abstract}
\section{Introduction}
In the theory and practice of cryptographic information protection, one of the key tasks
is the formation of PRS which width, length and characteristics meet modern
requirements~\cite{Kle1}. Many existing solutions in this area aim to obtain
a binary PRS of maximum memory length with acceptable statistical characteristics \cite{Sch2}.
However, recently it is considered that one of the further directions in the development of means
of information security~(MIS) is the use of multi-valued functions of the algebra of logic~(MFAL),
in particular, using the PRS over the Galois field GF($q$) ($q>2$), which have a wider spectrum
of unique properties comparing to binary PRS~\cite{Lid3}.

The nodes of the formation of the $q$-valued PRS, like the others, are prone to failures
and malfunction, which leads to the occurrence of errors in their functioning. In addition
to random errors occurrence in the generation of PRS related to ‘‘unintentional’’ failures
and malfunctions caused by various causes: aging of the element base, environmental influences,
severe operating conditions, etc. (reasons typical for reliability theory), there are deliberate
actions of an attacker aimed to create massive failures of electronic components of the formation
nodes of PRS due to the hardware errors generation (one of the types of information security
threats)~\cite{Yan4}.

Many methods have been developed to provide the necessary level of reliability of the digital
devices functioning; the most common are backup methods and methods of noise-immune coding.
However, backup methods do not provide the necessary levels of operation reliability with
limitations on hardware costs, and methods of noise-immune coding are not fully adapted
to the specifics of the construction and operation of MIS, in particular, generators of
$q$-valued PRS.

The work \cite{Fin5} offers a solution that overcomes the complexity of using code control
for the nodes of the binary PRS generation, based on the ‘‘arithmetic’’ of logical count and
the application of the redundant modular code device, which provides the necessary level
of security for their functioning. However, the solution obtained is limited to exclusive
applicability in the formation of binary PRS. At the same time, work~\cite{Fin6}, is known
where by means of ‘‘arithmetic’’ of logical count the task of parallelizing the nodes of forming
of binary PRS is solved, but without monitoring their functioning. As a result, it becomes
necessary to generalize the solutions obtained to ensure the security of the functioning
of the nodes of $q$-valued PRS formation.
\section{General Principles of Building Generators of $q$-valued PRS}
The most common and tested methods for PRS are algorithms and devices of PRS generation~---
linear recurrent shift registers ($q$-LFSR) with feedback~--- based on the use of recurrent
logical expressions~\cite{Sch2}.

The construction of the $q$-LFSR over the field GF($q$) is carried out from the given generating
polynomial:
\begin{align}\label{1}
K(x)=\sum_{i=0}^m k_{m-i}x^{m-i},
\end{align}
where $m$~--- is the polynomial degree $K(x)$, $m\in N$; $k_{i}\in GF(q)$, $k_{m}=1$, $k_{0}\neq0$.

Thus, the $q$-LFSR element is formed in accordance with the following characteristic equation~\cite{Mac7}:
\begin{align}\label{2}
    a_{p+m}=-k_{m-1}a_{p+m-1}-k_{m-2}a_{p+m-2}-\ldots-k_1a_{p+1}-k_0a_p.
\end{align}

The Eq.~(\ref{2}) is a recursion which describes an infinite $q$-valued PRS with period $q^{m}-1$
(with nonzero initial state, as well as under condition that the polynomial (\ref{1}) is primitive over
the field GF($q$)), each nonzero state appears once per period.

A homogeneous recurrent Eq.~(\ref{2}) can be presented in the following form:
\begin{align*}
    a_{p+m}=k_{m-1}a_{p+m-1}\oplus k_{m-2}a_{p+m-2}\oplus \ldots\oplus k_1a_{p+1}\oplus k_0a_p
\end{align*}
or
\begin{align}\label{3}
    a_{p+m}=\bigoplus_{i=1}^{m}k_{i-1}a_{p+i-1},
\end{align}
where $\oplus$~--- is the symbol of addition on module~$q$.

The $q$-LFSR corresponding to the polynomial (\ref{3}) is shown in Fig.~1, whose cells contain
field GF($q$) elements: $a_p,\, \ldots,\, a_{p+m-1}$.

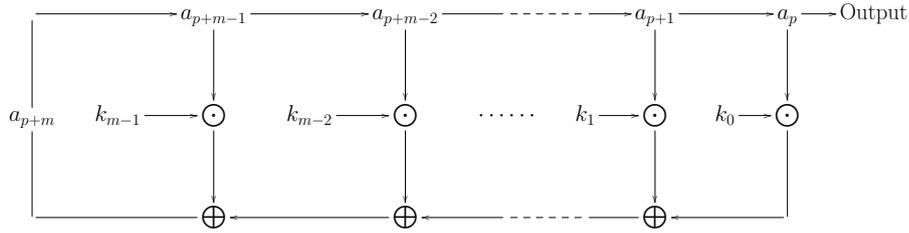
\begin{figure*}[!]
    \begin{center}\LARGE{
        \resizebox{1.0\linewidth}{!}{
$$
\xymatrix{
\ar@{->}[rr]\ar@{-}[dd] &          & a_{p+m-1}  \ar@{->}[dd] \ar@{->}[rr] &             & a_{p+m-2}\ar@{->}[dd] \ar@{-}[r]&  \ar@{--}[r]    &   \ar@{->}[r]         &  a_{p+1}   \ar@{->}[dd] \ar@{->}[rr]    &   &  a_{p}  \ar@{->}[dd] \ar@{->}[r]  & \mathrm{Output}\\
&&&&&&&&&&\\
a_{p+m}\ar@{-}[dd] &k_{m-1} \ar@{->}[r] & \bigodot \ar@{->}[dd]  &k_{m-2} \ar@{->}[r] & \bigodot \ar@{->}[dd]&  \cdots\cdots   &k_{1}  \ar@{->}[r]  &\bigodot \ar@{->}[dd]  & k_{0}  \ar@{->}[r] &\bigodot \ar@{-}[dd]                                    & \\
 &&&&&&&&&&\\
                                      &                   &  \bigoplus \ar@{-}[ll]                        &                                  & \ar@{->}[ll]\bigoplus    &   \ar@{->}[l]               &   \ar@{--}[l]                            &\ar@{-}[l] \bigoplus &                              & \ar@{->}[ll]                            &\\
 }
$$
}}
  \end{center}
             \caption{Structural diagram of the operation of the sequential  $q$-LFSR in accordance with formula (3)  ($\oplus$ and  $\odot$~--- according to transaction of addition and multiplication of the $\mod q$)}
        \end{figure*}
%
\section{Analysis of Possible Modifications $q$-valued PRS Caused by the Error Occurred}
It is known that the consequences of accidental errors that occur during the PRS generation
associated with ‘‘unintentional’’ failures, as well as the consequences of intentional
actions by an attacker based on the use of thermal, high-frequency, ionizing or other external
influences in order to obtain mass malfunctions of the equipment by initiation of calculation errors,
lead to similar types of PRS modification.

\begin{figure}[ht]
    \begin{center}
        \includegraphics[scale=.7]{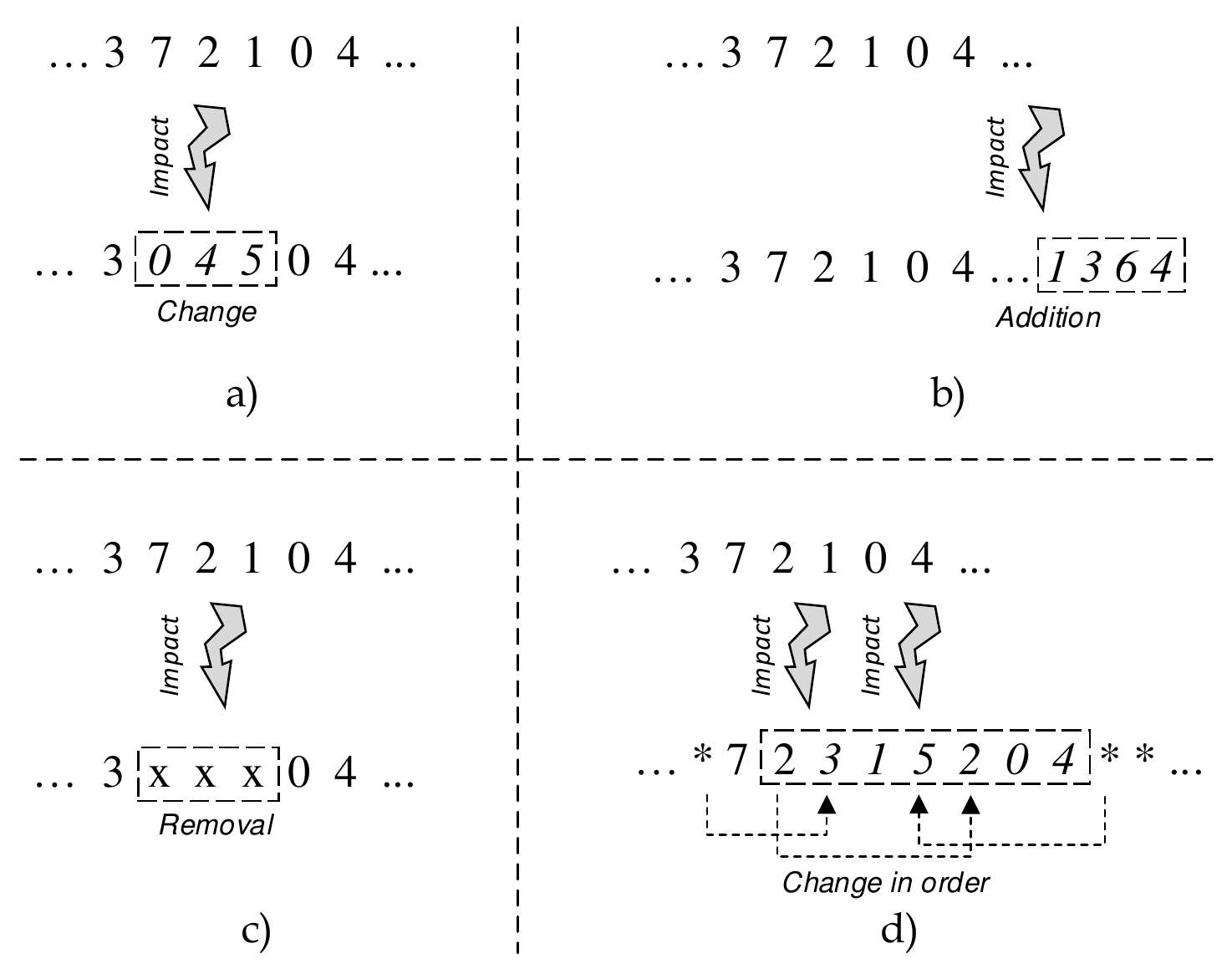}\end{center}
        \caption{The main types of PRS modification:
    a) change in the elements of the PRS,
    b) addition of new PRS elements,
    c) removal of the CAP elements,
    d) change in the order of the PRS elements}
\end{figure}

Fig.~2 shows main types of modification of PRS over the GF($q$) field.
The attacker’s actions based on error generation are highly effective for most of the known and
currently used algorithms for generating $q$-valued PRS~\cite{Can8,Car9,Pag10}. It is
known~\cite{Gut11} that the probability of error generation is proportional to the irradiation
time of the respective registers in a favorable state for the error occurrence and to the number
of bits within which an error is expected. This type of impact has not been sufficiently studied
and therefore represents a threat to the information security of modern and promising MIS functioning.

One of the ways to solve this problem is to develop a technique for improving the safety of the operation
of the MIS nodes most susceptible to these effects, in particular, the nodes of $q$-valued PRS formation.
\section{Analysis of Ways to Control the Generation of $q$-valued PRS}
Currently, the necessary level of security for the functioning of the nodes for the $q$-valued PRS formation
is achieved both through the use of redundant equipment (structural backup) and temporary redundancy due
to various calculations repetition.

In the field of digital circuit design solutions based on the use of block redundant coding methods are known.
To apply these methods to $q$-valued PRS generators it is necessary to solve the problem of parallelizing
the calculation process of the $q$-valued PRS.

The solution of the problem is based on the use of classical parallel recursion calculation algorithms~\cite{Ort12},
for which the characteristic Eq.~(\ref{3}) corresponding to the generating polynomial (\ref{2})
can be represented as a system of characteristic equations:
\begin{align}\label{4}
    \begin{cases}
        a_{t,\, m-1}=\bigoplus\limits_{i=1}^{m}k_{i-1}^{(m-1)}a_{t-1,\, p+i-1},\\
        a_{t,\, m-2}=\bigoplus\limits_{i=1}^{m}k_{i-1}^{(m-2)}a_{t-1,\, p+i-1},\\
        \cdots\cdots\cdots\cdots\cdots\cdots\cdots\cdots\cdots\cdots\\
        a_{t,\, 1}=\bigoplus\limits_{i=1}^{m}k_{i-1}^{(1)}a_{t-1,\, p+i-1},\\
        a_{t,\, 0}=\bigoplus\limits_{i=1}^{m}k_{i-1}^{(0)}a_{t-1,\, p+i-1},
    \end{cases}
\end{align}
where $k_{i-1}^{(j)}\in$ GF($q$); $j=0,\, 1,\, \ldots,\, m-2,\, m-1$.

The system (\ref{4}) forms an information matrix:
\begin{align*}\mathbf{G}_{\textbf{Inf}}=
    \begin{Vmatrix}
        \ k^{(m-1)}_{0} & k^{(m-1)}_{1} & \ \ldots \ & k^{(m-1)}_{m-2} & k^{(m-1)}_{m-1} \ \\
        \ k^{(m-2)}_{0} & k^{(m-2)}_{1} & \ \ldots \ & k^{(m-2)}_{m-2} & k^{(m-2)}_{m-1} \ \\
        \ \vdots        & \vdots        & \ \ddots \ & \vdots          & \vdots \          \\
        \ k^{(1)}_{0}   & k^{(1)}_{1}   & \ \ldots \ & k^{(1)}_{m-2}   & k^{(1)}_{m-1} \   \\
        \ k^{(0)}_{0}   & k^{(0)}_{1}   & \ \ldots \ & k^{(0)}_{m-2}   & k^{(0)}_{m-1} \
    \end{Vmatrix}.
\end{align*}

Similar result can be obtained in another convenient way \cite{Kle1}:
\begin{align*}\mathbf{G}_{\textbf{Inf}}=
    \begin{Vmatrix}
        \ k_{m-1} & k_{m-2} & \ldots & \ k_1 \ & \ k_0 \ \\
            \ 1       & 0       & \ldots & \ 0 \   & \ 0 \   \\
            \ 0       & 1       & \ddots & \ 0 \   & \ 0 \   \\
            \ 0       & 0       & \ldots & \ 0 \   & \ 0 \   \\
            \ 0       & 0       & \ldots & \ 1 \   & \ 0 \   \\
    \end{Vmatrix}^{m},
\end{align*}
where the elements raised to the power $m$ are of a matrix which is created according to the known
rules of linear algebra for the calculation of the next $q$-valued element of the PRS $a_{p+m}$:
 \begin{gather*}
  \begin{Vmatrix}
     \ a_{p+m} \ \\ \ a_{p+m-1} \ \\ \ \vdots \ \\ \ a_{p+2} \ \\ \ a_{p+1} \ \\
   \end{Vmatrix}= \left|
      \begin{array}{c}
         \begin{Vmatrix}
     \ a_{p+m-1} \ \\ \ a_{p+m-2} \ \\ \ \vdots \ \\ \ a_{p+1} \ \\ \ a_{p} \ \\
   \end{Vmatrix} \cdot
   \begin{Vmatrix}
            \ k_{m-1} &  \ldots &  k_0  \\
            \ 1       &  \ldots &  0    \\
            \ 0       &  \ldots &  0    \\
            \ 0       &  \ldots &  0    \\
            \ 0       &  \ldots &  0    \\
\end{Vmatrix} \\
      \end{array}
    \right|_{q},
\end{gather*}
where $\left|\cdot\right|_{q}$~--- is the smallest nonnegative deduction of the number ‘‘$\cdot$’’ on module~$q$.

The technique for raising a matrix to the power can be performed with help of symbolic calculations
in any computer algebra system with the subsequent simplification (in accordance with the axioms
of the algebra and logic) of the elements of the resulting matrix of the form $Yk_{j}^{b}=k_{j}$
according to the rules: 1) $k_{j}^{b}=k_{j}$; 2) $Y=0$, for even $Y$ and $Y=1$, for odd $Y$.
Thus, we obtain the $t$-block of PRS:
\begin{align*}
    \mathbf{A}_{t}=\left|\mathbf{G}_{\textbf{Inf}}\cdot \mathbf{A}_{t-1}\right|_{q},
\end{align*}
where
\begin{align*}
\mathbf{A}_{t}&=
    \left[
\begin{array}{ccccc}
  a_{t,\, p+m-1} & a_{t,\, p+m-2} & \ldots & a_{t,\, 1} & a_{t,\, 0}
\end{array}
\right]^{\top}, \\
\mathbf{A}_{t-1}&=
    \left[
\begin{array}{ccccc}
        a_{t-1,\, p+m-1}& a_{t-1,\, p+m-2} & \ldots & a_{t-1,\, 1} & a_{t-1,\, 0}
    \end{array}
\right]^{\top}.
\end{align*}

To create conditions for the use of a separable linear redundant code, we obtain a generating matrix
$\mathbf{G}_{\textbf{Gen}}$, consisting of the information and verification matrixes by adding in the
(\ref{4}) test expressions:
\begin{align*}
    \begin{cases}
        a_{t,\, p+m-1}=\bigoplus\limits_{i=1}^{m}k_{i-1}^{(m-1)}a_{t-1,\, p+i-1},\\
        \cdots\cdots\cdots\cdots\cdots\cdots\cdots\cdots\cdots\cdots\\
        a_{t,\, 0}=\bigoplus\limits_{i=1}^{m}k_{i-1}^{(0)}a_{t-1,\, p+i-1},\\
        a^{*}_{t,\, p+r-1}=\bigoplus\limits_{i=1}^{r}c_{i-1}^{(r-1)}a_{t-1,\, p+i-1},\\
        \cdots\cdots\cdots\cdots\cdots\cdots\cdots\cdots\cdots\cdots\\
        a^{*}_{t,\, 0}=\bigoplus\limits_{i=1}^{r}c_{i-1}^{(0)}a_{t-1,\, p+i-1},
    \end{cases}
\end{align*}
where $k_{i-1}^{(j)},\, c_{i-1}^{(z)}\in$ GF($q$); $z=0,\, \ldots,\, r-1$; $r$~--- is the number of redundant
symbols of the applied linear code; $j=0,\, \ldots,\, m-1$.

The forming matrix takes the form:
\begin{align*}\mathbf{G}_{\textbf{Gen}}=
    \begin{Vmatrix}
        k^{(m-1)}_{0} & k^{(m-1)}_{1} & \ldots & k^{(m-1)}_{m-2} & k^{(m-1)}_{m-1} \\
        \vdots & \vdots & \ddots & \vdots & \vdots \\
        k^{(0)}_{0} & k^{(0)}_{1} & \ldots & k^{(0)}_{m-2} & k^{(0)}_{m-1} \\
        c^{(r-1)}_{0} & c^{(r-1)}_{1} & \ldots & c^{(r-1)}_{r-2} & c^{(r-1)}_{r-1} \\
        \vdots & \vdots & \ddots & \vdots & \vdots \\
        c^{(0)}_{0} & c^{(0)}_{1} & \ldots & c^{(0)}_{r-2} & c^{(0)}_{r-1}
    \end{Vmatrix}.
\end{align*}

Then the $t$-block of the $q$-valued PRS with test digits (linear code block)
\begin{align*}\textbf{A}^{*}_{t}=
    \begin{bmatrix}
        a_{t,\, p+m-1} & \ldots & a_{t,\, 0} & a^{*}_{t,\, p+r-1} & \ldots & a^{*}_{t,\, 0}
    \end{bmatrix}^{\top}
\end{align*}
is calculated as:
\begin{align*}
    \mathbf{A}^{*}_{t}=\left|\mathbf{G}_{\textbf{Gen}}\cdot \mathbf{A}_{t-1}\right|_{q}.
\end{align*}

The anti-jamming decoding procedure is performed using known rules \cite{Ham13}.

The use of linear redundant codes and ‘‘hot’’ backup methods is not the only option for
realizing functional diagnostics and increasing the fault tolerance of digital devices.
Important advantages for these purposes are found in arithmetic redundant codes, in particular,
the so-called AN-codes and codes of modular arithmetic (MA). However, arithmetic redundant
codes are not applicable to logical data types. In logical calculations, their structure collapses,
which leads to the impossibility of monitoring errors in logical calculations.

The use of arithmetic redundant codes to control logical data types must be ensured by the
introduction of additional procedures related to the ‘‘arithmetic’’ of the logical count.
\section{The Procedure for Parallelizing the Generation of $q$-valued PRS by Means of Arithmetic Polynomials}
Parallelizing the ‘‘calculation’’ processes of complex systems or minimizing the number of operations
involving the use of all resources makes it possible to achieve any utmost characteristic or quality
index, which in turn is necessary in most practically important cases. In turn, the new direction formed
at the end of the last century~-- parallel-logical calculations through arithmetic (numerical) polynomials
\cite{Mal14}, also allowed to provide ‘‘useful’’ structural properties. It became possible to use arithmetic
redundant codes to control logical data types and increase the fault tolerance of implementing devices by
representing arithmetic expressions \cite{Mal14} as logical operations, in particular, by linear numerical
polynomials (LNP) and their modular forms \cite{Fin15}.

In~\cite{Fin5} an algorithm for parallelizing the generation of binary PRS is presented based on the
representation of systems of generating recurring logical formulas by means of LNP offered by V.~D. Malyugin,
which allowed using the redundant modular code device to control the errors of the functioning of the PRS
generation nodes and, ensure the required safety of their functioning in the MIS.

To ensure the possibility of applying code control methods to generators of $q$-valued PRS, it is necessary
to solve the problem of parallelizing the process of calculating them, while in \cite{Fin6} in general terms,
approach for the synthesis of parallel generators of $q$-valued PRS on arithmetic polynomials is presented,
the essence of which is the following.

Let $a_0,\, a_1,\, a_2,\, \ldots,\, a_{m-1}, \ldots$~--- be the elements of the $q$-valued PRS satisfying the
recurrence Eq.~(\ref{3}). Knowing that random element $a_p~(p\geq m)$ of the sequence
$a_0,\, a_1,\, a_2,\, \ldots,\, a_{m-1},\, \ldots$ is determined by the preceding $m$ elements, let us present
the elements $a_{p+m},\, a_{p+m+1},\, \ldots,\, a_{p+2m-1}$ of the section of the $q$-valued PRS by the
length $m$ in the form of a system of characteristic equations:
\begin{gather}\label{5}
\begin{cases}
a_{p+m}=\bigoplus\limits_{i=1}^{m}k_{i-1}a_{ p+i-1}, \\
a_{p+m+1}=\bigoplus\limits_{i=1}^{m}k_{i-1}a_{ p+i}, \\
\ldots\ldots\ldots\ldots\ldots\ldots\ldots\ldots\ldots\\
a_{p+2m-1}=\bigoplus\limits_{i=1}^{m}k_{i-1}a_{ p+i+m-2},
\end{cases}
\end{gather}
where $\left[a_{p+m}\ a_{p+m+1} \ \ldots \ a_{p+2m-1}\right]$~--- is the vector of the $m$-state of the $q$-valued
PRS (or the internal state of the $q$-LFSR on $m$-cycle of work).

By analogy with \cite{Fin5} let us express the right-hand sides of the system (\ref{5}) through the given
initial conditions and let us write it as the $m$ MFAL system of $m$ variables:
\begin{align}\label{6}
    \begin{cases}
     f_{1}\left(a_p,\, a_{p+1},\, \ldots,\, a_{p+m-1}\right)=\bigoplus\limits_{i=1}^{m}k_{i-1}^{(0)}a_{ p+i-1}, \\
     f_{2}\left(a_p,\, a_{p+1},\, \ldots,\, a_{p+m-1}\right)=\bigoplus\limits_{i=1}^{m}k_{i-1}^{(1)}a_{ p+i-1}, \\
     \ldots\ldots\ldots\ldots\ldots\ldots\ldots\ldots\ldots\ldots\ldots\ldots\ldots\ldots\ldots \\
     f_{m}\left(a_p,\, a_{p+1},\, \ldots,\, a_{p+m-1}\right)=\bigoplus\limits_{i=1}^{m}k_{i-1}^{(m-1)}a_{ p+i-1},
\end{cases}
\end{align}
where the coefficients $k_{i-1}^{(j)}\in \{0,\, 1,\, \ldots,\, q-1\}$ ($i=1,\, \ldots,\, m$; $j=0,\, \ldots,\, m-1$)
are formed after expressing the right-hand parts of the system (\ref{5}) through given initial conditions.

It is known that random MFAL can be represented in the form of an arithmetic polynomial in simple
way \cite{Fin16,Kuk17}:
\begin{align}\label{7}
  L\left(a_p,\, a_{p+1},\, \ldots,\, a_{p+m-1}\right)=\sum_{i=0}^{q^{m-1}-1}l_{i}\ a_{p}^{i_0}a_{p+1}^{i_1}\ \ldots\ a_{p+m-1}^{i_{m-1}},
\end{align}
where $a_u \in \{0,\,1,\, \ldots,\, q-1\}$;
$u=0,\, \ldots,\, m-1$; $l_i$~--- $i$-coefficient of an arithmetic polynomial;
$\left(i_0 \ i_1 \ \ldots \ i_{m-1}\right)_q$~--- representation of the parameter $i$ in the $q$-scale of notation:
\begin{align*}
\left(i_0 \ i_1 \ \ldots \ i_{m-1}\right)_q&=\sum_{u=0}^{m-1}i_{u}q^{m-u-1}\ \ \ (i_{u}\in{0,\, 1,\, \ldots,\, q-1});\\
 a_{u}^{i_{u}}&=
\begin{cases}
    1,& i_u=0,\\
    a_u,& i_u\neq 0.
\end{cases}
\end{align*}

Similar to \cite{Fin16,Kuk17} let us implement the MFAL system (\ref{6}) by computing some arithmetic polynomial.
In order to do this, we associate the MFAL system (\ref{6}) with a system of arithmetic polynomials of the
form (\ref{7}), we obtain:
   \begin{align}\label{8}
    \begin{cases}
     L_1\left(a_p,\, a_{p+1},\, \ldots,\, a_{p+m-1}\right)=\sum\limits_{i=0}^{q^{m-1}-1}l_{1,\, i}\ a_{p}^{i_0}a_{p+1}^{i_1}\ \ldots\ a_{p+m-1}^{i_{m-1}},\\
     L_2\left(a_p,\, a_{p+1},\, \ldots,\, a_{p+m-1}\right)=\sum\limits_{i=0}^{q^{m-1}-1}l_{2,\, i}\ a_{p}^{i_0}a_{p+1}^{i_1}\ \ldots\ a_{p+m-1}^{i_{m-1}},\\
      \ldots\ldots\ldots\ldots\ldots\ldots\ldots\ldots\ldots\ldots\ldots\ldots\ldots\ldots\ldots\ldots\ldots\ldots\ldots\ldots\\
       L_m\left(a_p,\, a_{p+1},\, \ldots,\, a_{p+m-1}\right)=\sum\limits_{i=0}^{q^{m-1}-1}l_{m,\, i}\ a_{p}^{i_0}a_{p+1}^{i_1}\ \ldots\ a_{p+m-1}^{i_{m-1}}.
      \end{cases}
      \end{align}

Let us multiply the polynomials of the system (\ref{8}) by weights  $q^{e-1}$ $(e=1,\, 2,\, \ldots,\, m)$:
\begin{align*}
\begin{cases}
     L_1^*\left(a_p,\, a_{p+1},\, \ldots,\, a_{p+m-1}\right)=q^{0}L_{1}\left(a_p,\, a_{p+1},\, \ldots,\, a_{p+m-1}\right)\\=\sum\limits_{i=0}^{q^{m-1}-1}l_{1, i}^{*}\ a_{p}^{i_0}a_{p+1}^{i_1}\ \ldots\ a_{p+m-1}^{i_{m-1}}, \\
     L_2^*\left(a_p,\, a_{p+1},\, \ldots,\, a_{p+m-1}\right)=q^{1}L_{2}\left(a_p,\, a_{p+1},\, \ldots,\, a_{p+m-1}\right)\\=\sum\limits_{i=0}^{q^{m-1}-1}l_{2, i}^{*}\ a_{p}^{i_0}a_{p+1}^{i_1}\ \ldots\ a_{p+m-1}^{i_{m-1}},\\
      \ldots\ldots\ldots\ldots\ldots\ldots\ldots\ldots\ldots\ldots\ldots\ldots\ldots\ldots\ldots\ldots\ldots\ldots\ldots\ldots\\
      L_m^*\left(a_p,\, a_{p+1},\, \ldots,\, a_{p+m-1}\right)=q^{m-1}L_{m}\left(a_p,\, a_{p+1},\, \ldots,\, a_{p+m-1}\right)\\=\sum\limits_{i=0}^{q^{m-1}-1}l_{m, i}^{*}\ a_{p}^{i_0}a_{p+1}^{i_1}\ \ldots\ a_{p+m-1}^{i_{m-1}},
       \end{cases}
       \end{align*}
where $l^{*}_{e,\, i}=q^{e-1}l_{e,\, i}$  $(e=1, 2, \ldots, m;\ \ \ i=0, \ldots, q^m-1)$.

Then we get:
 \begin{align}\label{9}
    L\left(a_p,\, a_{p+1},\, \ldots,\, a_{p+m-1}\right)=\sum_{i=0}^{q^{m-1}-1}\sum_{e=1}^{d}l^{*}_{e,\, i}\ a_{p}^{i_0}a_{p+1}^{i_1}\ \ldots\ a_{p+m-1}^{i_{m-1}}
       \end{align}
or using the provisions of \cite{Asl18}:
\begin{align}\label{10}
    D\left(a_p,\, a_{p+1},\, \ldots,\, a_{p+m-1}\right)=\left|\bigoplus_{i=0}^{q^{m-1}-1}v_{i}\ a_{p}^{i_0}a_{p+1}^{i_1}\ \ldots\ a_{p+m-1}^{i_{m-1}}\right|_{q^m},
\end{align}
where $$v_i=\bigoplus_{e=1}^{m}l^{*}_{e,\, i}\  (i=0,\, 1,\, \ldots,\, q^{m-1}-1).$$

Let us calculate the values of the desired MFAL. For this, the result of the calculation (\ref{10}) is
presented in the $q$-scale of notation and we apply the camouflage operator $\Xi^w\{D\left(a_p,\, a_{p+1},\, \ldots,\, a_{p+m-1}\right)\}$:

  $$\Xi^w\{D\left(a_p,\, a_{p+1},\, \ldots,\, a_{p+m-1}\right)\}=\left|\left \lfloor \frac{D\left(a_p,\, a_{p+1},\, \ldots,\, a_{p+m-1}\right)}{q^w} \right \rfloor\right|_{q}$$,
where $w$~--- is the desired $q$-digit of the representation $D\left(a_p,\, a_{p+1},\, \ldots,\, a_{p+m-1}\right)$.

The presented method, based on the MFAL arithmetic representation, makes it possible to control
the $q$-valued PRS generation errors by means of arithmetic redundant codes.
\section{Control of Errors in the Operation of Generators of $q$-valued PRS by Redundant MA Codes}
In MA, the integral nonnegative coefficient $l^{*}_{e,\, i}$ of an arithmetic polynomial (\ref{9})
is uniquely presented by a set of balances on the base of MA
($s_{1},\, s_{2},\, \ldots,\, s_{\eta}< \\< s_{\eta+1}<\ldots<s_{\psi}$~--- simple pairwise):
\begin{align}\label{11}
    l^{*}_{e,\, i}=(\alpha_{1},\, \alpha_{2},\, \ldots,\, \alpha_{\eta},\, \alpha_{\eta+1},\, \ldots,\, \alpha_{\psi})_{\text{MA}},
\end{align}
where $\alpha_{\tau}=\left|l^{*}_{e,\, i}\right|_{s_{\tau}}$; $\tau=1,\, 2,\, \ldots,\, \eta,\, \ldots,\, \psi$.
The working range $S_{\eta}=s_{1}s_{2}\ldots s_{\eta}$ must satisfy $S_{\eta}>2^{g}$, where
$g=\sum\limits_{1\leq\varepsilon\leq\sigma}\theta_{\varepsilon}$~--- is the number of bits required
to represent the result of the calculation (\ref{9}).

Balances $\alpha_{1},\, \alpha_{2},\, \ldots,\, \alpha_{\eta}$ are informational, and
$\alpha_{\eta+1},\, \ldots,\, \alpha_{\psi}$~--- are control. In this case, MA is called extended and
covers the complete set of states presented by all the $\psi$ balances. This area is the full MA range
$[0,\, S_{\psi})$, where $S_{\psi}=s_{1}s_{2}\ldots s_{\eta}s_{\eta+1}\ldots s_{\psi}$, and consists
of the operating range $[0,\, S_{\eta})$, defined by the information bases of the MA, and the range
defined by the redundant bases $[S_{\eta},\, S_{\psi})$, representing an invalid area for the results
of the calculations. This means that operations on numbers $l^{*}_{e,\, i}$ are performed in the range
$[0,\, S_{\psi})$. Therefore, if the result of the MA operation goes beyond the limits $S_{\eta}$, then
the conclusion about the calculation error follows.

Let us study the MA given by the $s_{1},\, s_{2},\, \ldots,\, s_{\eta},\, \ldots,\, s_{\psi}$ bases.
Each coefficient $l^{*}_{e,\, i}$ of a polynomial (\ref{9}) is presented in the form (\ref{11})
and we obtain an MA redundant code, represented by a system of polynomials:
\begin{align}\label{12}
   \begin{cases}
        U^{(1)}=L^{(1)}\left(a_p,\, a_{p+1},\, \ldots,\, a_{p+m-1}\right)=\sum\limits_{i=0}^{q^{m-1}-1}\sum_{e=1}^{d}l^{*(1)}_{e,\, i}\ a_{p}^{i_0}a_{p+1}^{i_1} \ldots a_{p+m-1}^{i_{m-1}}, \\
        U^{(2)}=L^{(2)}\left(a_p,\, a_{p+1},\, \ldots,\, a_{p+m-1}\right)=\sum\limits_{i=0}^{q^{m-1}-1}\sum_{e=1}^{d}l^{*(2)}_{e,\, i}\ a_{p}^{i_0}a_{p+1}^{i_1} \ldots a_{p+m-1}^{i_{m-1}}, \\
        \cdots\cdots\cdots\cdots\cdots\cdots\cdots\cdots\cdots\cdots\cdots\cdots\cdots\cdots\cdots\cdots\cdots\cdots\cdots\cdots\cdots\cdots\cdots\cdots\cdots\\
        U^{(\eta)}=L^{(\eta)}\left(a_p,\, a_{p+1},\, \ldots,\, a_{p+m-1}\right)=\sum\limits_{i=0}^{q^{m-1}-1}\sum_{e=1}^{d}l^{*(\eta)}_{e,\, i}\ a_{p}^{i_0}a_{p+1}^{i_1} \ldots a_{p+m-1}^{i_{m-1}}, \\
        \cdots\cdots\cdots\cdots\cdots\cdots\cdots\cdots\cdots\cdots\cdots\cdots\cdots\cdots\cdots\cdots\cdots\cdots\cdots\cdots\cdots\cdots\cdots\cdots\cdots\\
        U^{(\psi)}=L^{(\psi)}\left(a_p,\, a_{p+1},\, \ldots,\, a_{p+m-1}\right)=\sum\limits_{i=0}^{q^{m-1}-1}\sum_{e=1}^{d}l^{*(\psi)}_{e,\, i}\ a_{p}^{i_0}a_{p+1}^{i_1} \ldots a_{p+m-1}^{i_{m-1}}.
    \end{cases}
\end{align}

Substituting in (\ref{12}) the values of the MA balances for the corresponding bases for each coefficient
(\ref{9}) and the values of the variables $a_{p},\, a_{p+1},\, \ldots,\, a_{p+m-1}$, we obtain the values
of the polynomials of the system (\ref{12}), where $U^{(1)},\, U^{(2)},\, \ldots,\, U^{(\eta)},\, \ldots,\, U^{(\psi)}$~--- are
nonnegative integrals. In accordance with the Chinese balances theorem, we solve the system of equations:
\begin{align}\label{13}
\begin{cases}
    U^{*}=\left|U^{(1)}\right|_{s_{1}},\\
    U^{*}=\left|U^{(2)}\right|_{s_{2}},\\
    \ldots\ldots\ldots\ldots\\
    U^{*}=\left|U^{(\eta)}\right|_{s_{\eta}},\\
    \ldots\ldots\ldots\ldots\\
    U^{*}=\left|U^{(\psi)}\right|_{s_{\psi}}.
\end{cases}
\end{align}

Since $s_{1},\, s_{2},\, \ldots,\, s_{\eta},\, \ldots,\, s_{\psi}$ are simple pairwise, the only solution (\ref{13})
gives the expression:
\begin{align}\label{14}
    U^{*}=\left|\sum_{d=1}^{\psi}S_{d,\, \psi}\mu_{d,\, \psi}U^{(d)}\right|_{S_{\psi}},
\end{align}
where $S_{d,\, \psi}=\dfrac{S_{\psi}}{s_{d}}$, $\mu_{d,\, \psi}=\left|S^{-1}_{d,\, \psi}\right|_{s_{d}}$,
$S_{\psi}=\prod\limits_{d=1}^{\psi}s_{d}$.

The occurrence of the calculation result (\ref{14}) in the range (test expression)
\begin{align*}
    0\leq U^{*}<S_{\eta},
\end{align*}
means no detectable calculation errors.

Otherwise, the procedure for restoring the reliable functioning of the $q$-valued PRS
generator can be implemented according to known rules~\cite{Omo19}.
\section{Conclusion}
A secure parallel generator of $q$-valued PRS on arithmetic polynomials is presented.
The implementation of generators of $q$-valued PRS using arithmetic polynomials and redundant
MA codes makes it possible to obtain a new class of solutions aimed to safely implement
logical cryptographic functions. At the same time, both functional monitoring of equipment
(in real time, which is essential for~MIS) and its fault tolerance is ensured due to the possible
reconfiguration of the calculator structure in the process of its degradation. The classical
$q$-LFSR, studied in this work, forms the basis of more complex $q$-valued PRS generators.


\begin{thebibliography}{00} 
\bibitem{Kle1}
Klein,~A.: Stream Ciphers. Springer, http://www.springer.com. (2013)

\bibitem{Sch2}
Schneier,~B.: Applied Cryptography. Wiley, New York (1996)

\bibitem{Lid3}
Lidl,~R., Niederreiter,~H.: Introduction to finite fields and their applications. Cambridge: Cambridge Univ. Press. (1987)

\bibitem{Yan4}
Yang,~B., Wu,~K., Karri,~R.: Scan based side channel attack on data encryption standard. Report \textbf{2004}(324), 114--116 (2004)

\bibitem{Fin5}
Finko,~O.A., Dichenko,~S.A.: Secure Pseudo-Random Linear Binary Sequences Generators Based on Arithmetic Polynoms.
Advances in Intelligent Systems and Computing, Soft Computing in Computer and Information Science, \textbf{342}, Springer, Cham, pp.~279--290 (2015)

\bibitem{Fin6}
Finko,~O.A., Samoylenko,~D.V., Dichenko,~S.A., Eliseev,~N.I.: Parallel generator of $q$-valued pseudorandom sequences based on arithmetic polynomials.
Przeglad Elektrotechniczny, \textbf{3}, pp.~24--27 (2015)

\bibitem{Mac7}
MacWilliams,~F., Sloane,~N.: Pseudo-random sequences and arrays, Proc. IEEE, \textbf{64}, pp.~1715--1729 (1976)

\bibitem{Can8}
Canovas,~C., Clediere,~J.: What do DES S-boxes say in differential side channel attacks? Report \textbf{2005}(311), 191--200 (2005)

\bibitem{Car9}
Carlier,~V., Chabanne,~H., Dottax,~E.: Electromagnetic side channels of an FPGA implementation of AES. Report \textbf{2004}(145), 111--124 (2004)

\bibitem{Pag10}
Page,~D.: Partitioned cache architecture as a side-channel defence mechanism. Report \textbf{2005}(280), 213--225 (2005)

\bibitem{Gut11}
Gutmann,~P.: Software generation of random numbers for cryptographic purposes. Usenic security symp., usenix assoc., berkeley,
pp.~243--257, Calif (1998)

\bibitem{Ort12}
Ortega,~J.M.: Introduction to Parallel \& Vector Solution of Linear Systems. Plenum Press, New York (1988)

\bibitem{Ham13}
Hamming,~R.: Coding and Information Theory. Prentice-Hall (1980)

\bibitem{Mal14}
Malyugin,~V.D.: Representation of boolean functions as arithmetic polynomials. Autom. Remote. Control. \textbf{43}(4), 496--504 (1982)

\bibitem{Fin15}
Finko,~O.A.: Large systems of boolean functions: realization by modular arithmetic methods. Autom. Remote. Control. \textbf{65}(6),
871--892 (2004). June

\bibitem{Fin16}
Finko,~O.A.: Modular forms of systems of $k$-valued functions of the algebra of logic. Autom. Remote. Control. \textbf{66}(7), 1081--1100 (2005)

\bibitem{Kuk17}
Kukharev,~G.A., Shmerko,~V.P., Zaitseva,~E.N.: Algorithms and systolic processors of multivalued data. Minsk: Science and Technology (1990) (in Russian)

\bibitem{Asl18}
Aslanova,~N.H., Faradzhev,~R.G.: Arithmetic representation of functions of many-valued logic and parallel algorithm for finding such a representation.
Autom. Remote. Control. \textbf{53}(2), 251--261 (1992)

\bibitem{Omo19}
Omondi,~A., Premkumar,~B.: Residue Number System: Theory and Implementation. Imperial Collegt Press, London (2007)
\end{thebibliography}
\end{document}